\documentclass[reprint,
superscriptaddress,
 amsmath,amssymb,
 aps,
]{revtex4-2}

\usepackage{graphicx,float}
\usepackage{xcolor}
\usepackage{dcolumn}
\usepackage{bm}
\usepackage{cleveref}
\usepackage{siunitx}
\newcommand{\EuSnAs}{Eu\textsubscript{5}Sn\textsubscript{2}As\textsubscript{6}}

\usepackage{xr}

\begin{document}

\title{Colossal Magnetoresistance and Anisotropic Spin Dynamics in Antiferromagnetic Semiconductor Eu$_5$Sn$_2$As$_6$}

\author{R. P. \surname{Day}}
\thanks{These two authors contributed equally}
\affiliation{Department of Physics, University of California, Berkeley, California 94720, USA}

\author{K. \surname{Yamakawa}}
\thanks{These two authors contributed equally}
\email{kohtaro\_yamakawa@berkeley.edu}
\affiliation{Department of Physics, University of California, Berkeley, California 94720, USA}

\author{L. \surname{Pritchard Cairns}}%
\affiliation{Department of Physics, University of California, Berkeley, California 94720, USA}
\author{J. \surname{Singleton}}
\affiliation{Los Alamos National High Magnetic Field Lab}
\author{M. \surname{Allen}}%
\affiliation{Department of Physics, University of California, San Diego, CA 92093, USA}
\author{J. E. \surname{Moore}}%
\affiliation{Department of Physics, University of California, Berkeley, California 94720, USA}
\author{James G. Analytis}%
  \email{analytis@berkeley.edu}
\affiliation{University of California, Berkeley}
\affiliation{CIFAR Quantum Materials, CIFAR, Toronto, Canada}
\affiliation{Kavli Energy NanoScience Institute, Berkeley, CA, USA}

\date{\today}

\begin{abstract}
We report on the thermodynamic and transport properties of the rare-earth Zintl compound Eu$_5$Sn$_2$As$_6$, which orders as a canted antiferromagnetic magnetic semiconductor at 10.3~K. The system also displays a complex cascade of magnetic phases arising from geometric and magnetic exchange frustration, with high sensitivity to the application and direction of small magnetic fields.  At low temperature, Eu$_5$Sn$_2$As$_6$ exhibits negative colossal magnetoresistance of up to a factor of $6\times10^3$. This may be a lower bound as the conductivity appears to be shunted by an unknown conduction channel, causing the resistivity to saturate. Mechanisms for the low temperature saturation of resistivity are discussed.\end{abstract}

\maketitle


\section{Introduction}
Magnetic semiconductors present a unique opportunity to explore the interplay of disparate energy scales, giving rise to a diversity of electronic and magnetic phenomena \cite{haas_magnetic_1970}. The rare-earth Zintl compounds are of particular interest, which introduce local-moment magnetism, low carrier densities, narrow band gaps and dimensional confinement of both itinerant carriers and local moments. This combination is believed to give rise to magnetic topological insulators \cite{Rosa_2020, Varnava_2022}, exceptional thermoelectric material properties \cite{Wang_2012, Childs_2019}, Kondo lattice physics \cite{Goforth_2008} and unconventional superconductivity \cite{Ren_2009}, among others. 

The quasi-one dimensional \EuSnAs{} is one such example of the rare-earth Zintl semiconductors, belonging to a structural family that include antiferromagnetic topological insulators.\cite{Rosa_2020, Varnava_2022} In this work, we explore the low temperature electrical and magnetic properties of this system. We observe a hierarchy of orders as a function of temperature and magnetic field, operating across distinct length and energy scales. Beginning at temperatures well above that of long-range order, magnetic clusters begin to form, binding to the dilute free carriers in the system consistent with the formation of magnetic polarons. As temperature is reduced, this gives way to a cascade of metamagnetic transitions, as well as glassy dynamics in very low field. The interplay of these various phases within the local Eu$^{2+}$ ions and SnAs$_3$ chains is manifest in the form of colossal magnetoresistance spanning nearly four decades, reflecting the significant exchange interaction between local moments and itinerant carriers in this low-carrier density system. Finally, at low fields and temperatures, resistance appears to shunt across a low-impedance path. 

\section{Experimental Results}
\subsection{Methods}
\EuSnAs{} crystallizes in the orthorhombic space group $Pbam$ (No. 55), shown schematically in Fig~ \ref{fig:simple}(a). First synthesized as a candidate thermoelectric \cite{Wang_2012, Devlin_2022}, the structure can be described as composed of one-dimensional chains of corner-sharing Sn-As tetrahedra bound together by staggered plaquettes of Eu$^{2+}$ ions. Single crystals were prepared by a flux method. Europium metal (6N, Ames Lab) was cut up and combined with arsenic pieces (4N, Alfa Aesar) and tin shot (5N, Sigma Aldrich) in a 1:3:16 molar ratio in an alumina crucible under a dry nitrogen atmosphere. The crucible was sealed in an evacuated quartz ampoule without being exposed to air. The ampoule was heated in a furnace to \SI{400}{\degreeCelsius} over 2 hours, held for 1 hour, then heated to \SI{650}{\degreeCelsius} over 3 hours, held for 2 hours, then ramped to \SI{1000}{\degreeCelsius} over 10 hours before holding at \SI{1000}{\degreeCelsius} for 12 hours. Finally, the ampoule was slowly cooled to \SI{750}{\degreeCelsius} at \SI{2}{\degreeCelsius\per\hour}, and spun in a centrifuge to remove excess Sn flux. This process yields long bar-shaped crystals as illustrated in Fig.~\ref{fig:simple}(b) with a typical length of 3 - 10 mm and widths of 50 - 500 \unit{\micro\metre}.


\subsection{Magnetic Semiconductor with Canted Antiferromagnetism
}

\begin{figure}
    \centering
    \includegraphics[width=\columnwidth]{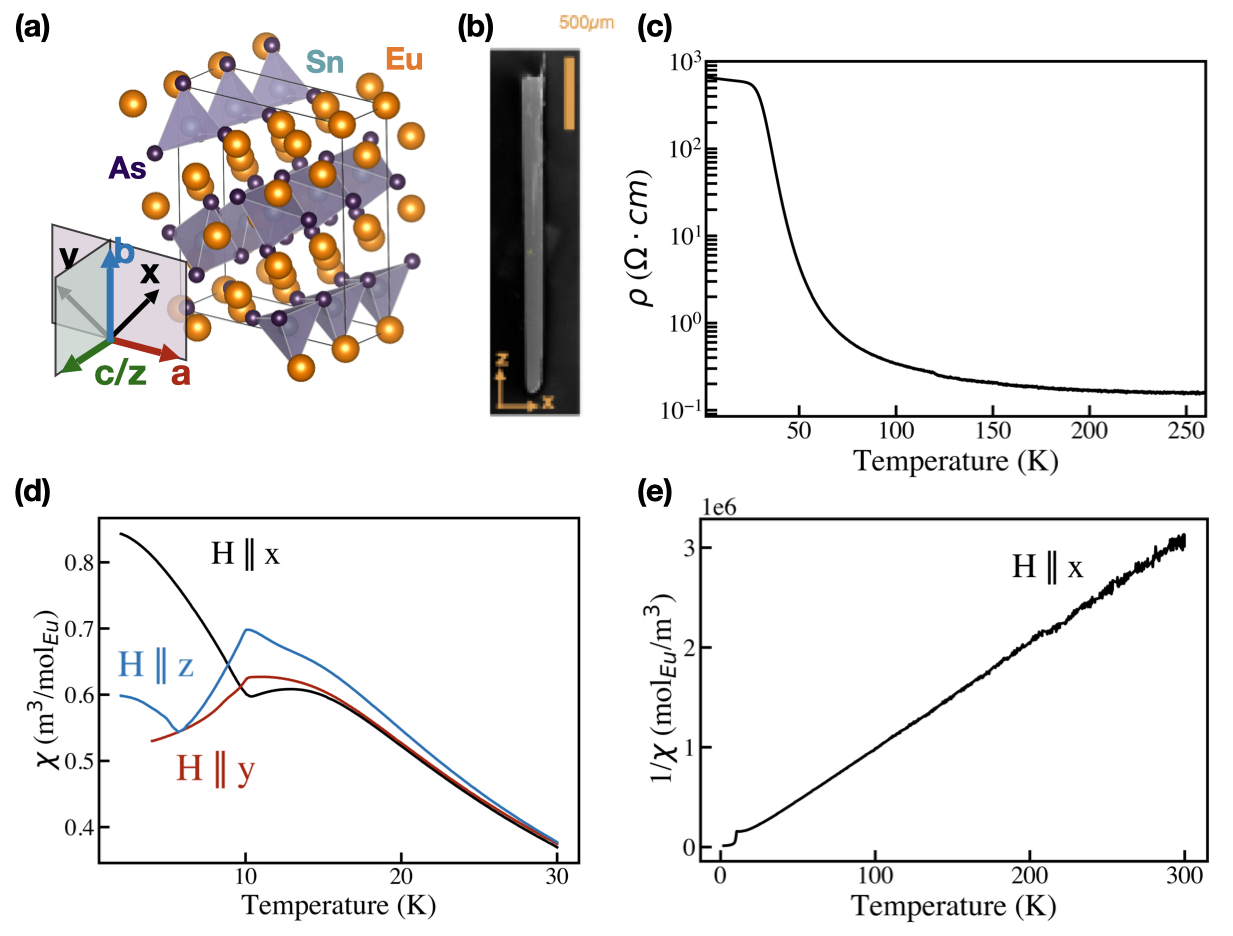}
    \caption{(a) The crystal structure of \EuSnAs{}, with the coordinate frame as used throughout the text. An SEM micrograph of a crystal is provided in (b), with the long axis along the Sn-As chains. In (c), a typical resistivity plot is shown, indicating a negative colossal magnetoresistance. The transition to canted AFM order visible below 10.3~K as seen in (d) where the magnetization under a bias field of 20~mT along $x$, $y$, and $z$ directions are shown in black, red, and blue respectively.  In (e), a plot of the inverted magnetic susceptibility under a 20~mT cooling field in the $z$ direction over the full temperature range is shown.}
    \label{fig:simple}
\end{figure}

The behaviour of \EuSnAs{} can be classified as that of a narrow-gap canted antiferromagnetic semiconductor. From a Curie-Weiss fit to the magnetic susceptibility - plotted in Fig.~\ref{fig:simple}(e) - the fluctuating moment is $\mu_\text{eff}=7.8(2)\mu_\text{B}/$Eu, indicating that Eu is in a 2+ valence, with a $J=S=7/2$ configuration. To within the experimental mass uncertainty of the crystal, this is in excellent agreement with the expectation of $7.93\mu_\text{B}$ for isolated Eu$^{2+}$ moments. Upon cooling, \EuSnAs{} acts similarly to an ordinary narrow gap semiconductor \cite{sze_semiconductor_2012}, as evidenced in Fig.~\ref{fig:simple}(c). At well above the onset of magnetic fluctuations, Hall effect measurements at 300K suggest a low hole-carrier density of \SI{1.75e18}{cm^{-3}}, indicating a very low concentration of impurity carriers. The low-field Hall number grows precipitously as the temperature is lowered (as shown in Fig.~\ref{fig:MR}(d)), suggesting the carrier density drops with a functional form typical of thermally activated carriers. However, the semiconducting nature of these systems and multi-band character \cite{Varnava_2022} makes a definitive determination of the true carrier density at these lower temperatures impossible from these measurements alone. Nevertheless, even as an approximation, the low carrier density of \EuSnAs{} contrasts with the majority of other Eu-based magnetic topological-insulator candidate materials, wherein the bulk carrier density is an order of magnitude larger~\cite{Sato_2020, Yan_2022,Li_2021}. This highlights why the study of this family of materials as magnetic topological insulators is needed.

In addition to the Hall result, low temperature resistance is well fitted to a simple activated Arrhenius function with a gap of \SI{35}{meV}, as plotted in Appendinx Fig.~\ref{app_fig:arrhenius}.
At the lowest temperatures, below $T\sim~20-30$~K, the resistance increases further only polynomially, resulting in a ``knee" in the data. We note that the precise temperature of this knee in $\rho_{zz}$(T) (along the long axis of the crystal, see Fig. \ref{fig:simple}(a)), as well as the level of saturation, is somewhat variable between samples. We discuss the origin of the saturation in Section~\ref{sec:discuss}.

\begin{figure}
    \centering
    \includegraphics[width=\columnwidth]{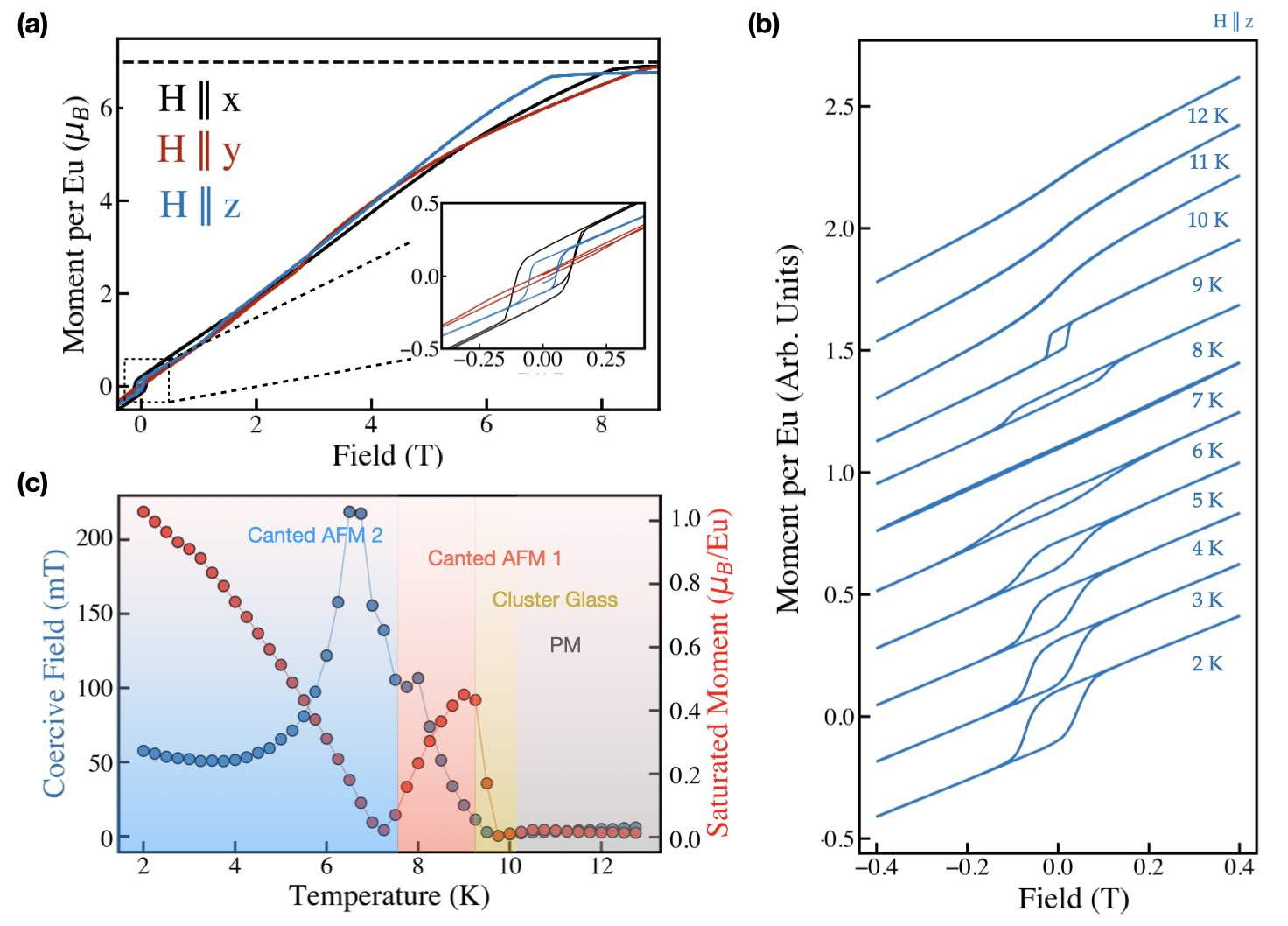}
    \caption{Magnetization measurements as a function of field. Isothermal field sweep plots are shown along the different field directions in panel (a), with the inset illustrating an enlargement of the low field region.  Hysteresis loops for $H\parallel z$ are plotted in (b) to show its evolution as a function of temperature. These loops are characterized in panel (c) with the saturation moment plotted in red and coercive field plotted in blue, with temperature regions labelled with the paramagnetic phase (PM), the cluster glass phase, and canted AFM phase 1 and 2.}
    \label{fig:phase}
\end{figure}

The magnetic susceptibility plotted in Fig.~\ref{fig:simple}(e) gives an initial impression of a ferromagnetic ground state -- on account of both the positive Curie-Weiss temperature $\Theta_\text{CW}=5.5$~K and the self-evident static magnetization below $10.3$~K. Indeed, a previous characterization of the material \cite{Wang_2012} reported \EuSnAs{} to be a ferromagnet. While we find our experimental results to be quantitatively consistent with this previous report, the presence of a minute static moment of approximately $0.2\mu_\text{B}$ per Eu (see hysteresis loops in Fig.~\ref{fig:phase}(a)), as opposed to a much smaller than the expected 7$\mu_\text{B}$ for Eu$^{2+}$, rule out this possibility. However, consideration of the magnetization when probed along different field-orientations cause us to classify this system as a semiconducting, canted antferromagnet. In Fig.~\ref{fig:simple}(d), we observe a suppression of the moment below $T_N$ when the field is along either the $y$ or $z$ directions. For fields oriented otherwise in the $x-y$ plane, static magnetization is observed for nearly all field projections away from $H\parallel y$. Such a strong angular anisotropy is somewhat surprising on account of the modest orthorhombicity of the material, but is otherwise corroborated via angle-dependent magnetoresistance shown in Appendix Fig.~\ref{fig:anisotropy}. The absence of a substantive net moment for $H\parallel y$  is made more clear through field-sweeps at fixed temperatures, plotted in Fig.~\ref{fig:phase}(a), where a small hysteresis loop is really only observed for fields along $x$ or $z$ but not $y$. 
\begin{figure*}
    \centering
    \includegraphics[width=\textwidth]{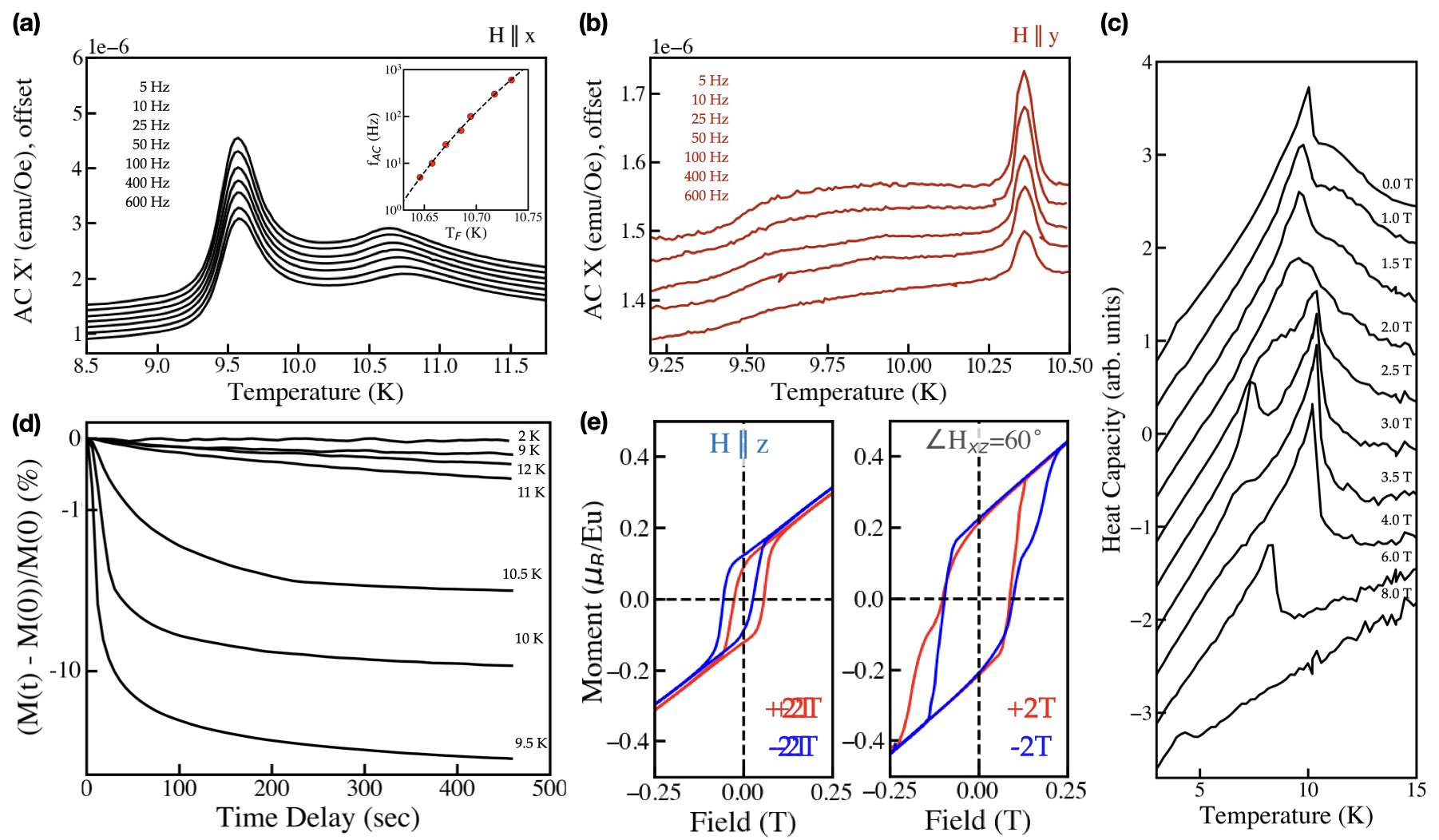}
  \caption{ Exchange competition in \EuSnAs{}.  Panels (a) and (b) plot frequency-dependent ACMS cooling curves for two sample orientations, with the inset in (a) illustrating the frequency-dependent shift of the higher-transition temperature $T_f$ and its corresponding Vogel-Fulcher fit.  In (c), heat capacity is plotted with applied field along the x direction indicating the splitting of two transitions near 10K.  In (d),  isothermal relaxation curves plots are shown, in which the sample is cooled to the specified temperature in an applied field of 0.1T along $z$. The field is removed and the magnetization of the sample is recorded as a function of time. The magnetization relaxes logarithmically only for $T_{N2} < T < T_{N1}$, indicating the presence of re-entrant long-range order below $T_{N2}$.  The exchange bias in the
  hysteresis loops for field-cooled samples oriented along $z$ and 60$^o$ off of $z$ are presented in (e), where the latter seems to go through a hysteretic metamagnetic transition.}
    \label{fig:magnetism}
\end{figure*}

\subsection{Competing orders, Metamagnetic transitions and Anisotropic Spin Dynamics}
\begin{figure*}
    \includegraphics[width=\textwidth]{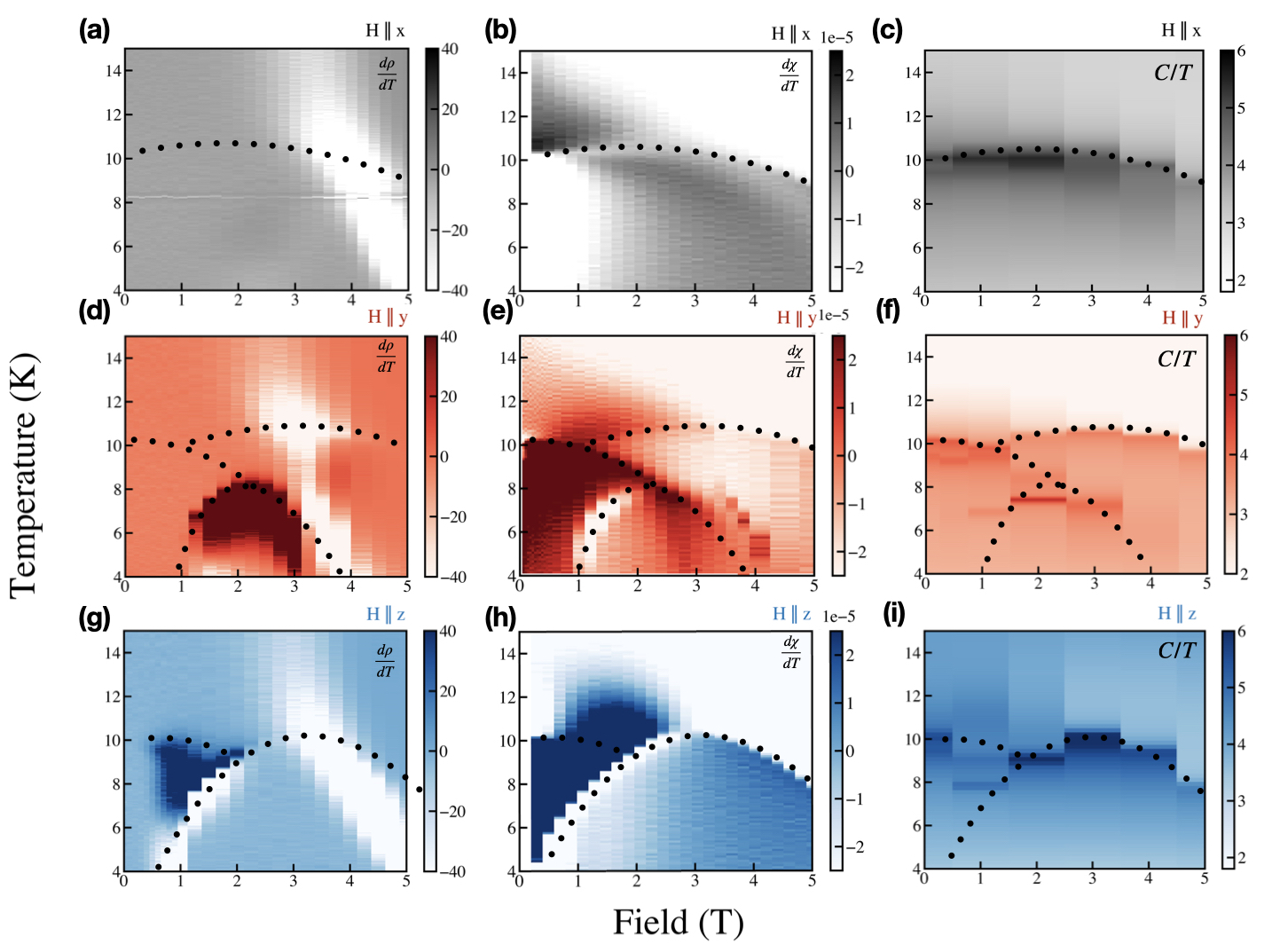}
    \caption{For fields projected along $x$, $y$, and $z$ directions, color diagrams of the temperature derivatives $d\rho/dT$ of the magnetoresistance (a-c)  temperature derivatives of the magnetic susceptibility $d\chi/dT$ (d-f), and heat capacity over temperature (g-i) are shown.  Dotted lines are drawn as guides to the eye, indicating how the magnetic phase transitions are captured through different characterization techniques.}
    \label{fig:colormap}
\end{figure*}

 While the broad characteristics of the magnetization do not reflect significant anisotropy (Fig.~\ref{fig:phase}(a))), for example through the absence of an easy axis or plane, a cascade of metamagnetic transitions are observed below 11 K with significant anisotropy.  Save for the outermost phase boundary which separates the ordered phase from the paramagnetic state, the intervening phases depend acutely on the orientation of the field. This observation is carried across all bulk probes of the material surveyed, including magnetic susceptibility $\chi_{ii}(H,T)$, heat capacity (Fig.~\ref{fig:magnetism}(c)), and resistivity $\rho_{ii}(H,T)$, as illustrated by the false colourplots in Fig. \ref{fig:colormap}. Their ubiquity across distinct physical probes, in addition to the sharp peaks in the magnetic heat capacity identify these transitions as metamagnetic phase transitions.
 


To best understand the nature of the phase transitions observed here, and the hierarchy of competing interactions in the material, we need to look more carefully at the magnetic lattice structure. Eu occupies three symmetry inequivalent sites in the lattice (Appendix Fig.~\ref{app_fig:hall}), and comparison of bond-lengths indicates that by considering direct exchange between neighbouring Eu-sites, there are at least 5 different exchange pathways to account for. Incorporating indirect-exchange mediated by the SnAs chains, or second-order Eu-Eu exchange at greater distances, only compounds this complexity. This observation is corroborated by the small, positive Curie Weiss temperature $\Theta_\text{CW}=5.5$~K, which contrasts with both the ordering temperature of $T_\text{N}=10.3$~K and the AFM-nature of the order. This notion of competing orders is further exemplified through close consideration of the evolution of the minor hysteresis loops plotted in Fig.~\ref{fig:phase}(c). 

Qualitatively, it is apparent that we do not observe growth of a single order parameter upon cooling below $T_\text{N}$, as the hysteresis loop changes in both saturation moment and coercive field in a non-monotonic fashion. This evolution is made quantitative in Fig.~\ref{fig:phase}(b) where we observe the collapse of the saturated moment around 7.4~K, followed by a recovery and continuous growth to low temperature. By contrast the coercive field of the hysteresis loops diverges around this collapse, before itself dropping to an approximately saturated low-temperature value. This gives a direct indication of competing order parameters in the system, with a static moment evolving below 7.4~K which is distinct from that of the original ordering at $T_\text{N}=10.3$~K. 

Further evidence to the presence of competing orders is found in the observation of exchange bias in the minor hysteresis loops, plotted in Fig.~\ref{fig:magnetism}(e) for two different field orientations. While exchange bias phenomena is perhaps most familiar from the context of AFM-FM heterostructures \cite{Stamps_2000}, similar physics is found in various systems with interacting or competing order parameters \cite{Maniv_2021}. 

We find that AC susceptibility measurements are helpful in elucidating the nature of these competing orders. In contrast to the vibrating sample magnetometry (VSM), where a magnetized sample oscillates through a co-axial pickup coil in the presence of a static DC bias field, the AC Magnetic Susceptibility (ACMS) introduces an additional coilset to the apparatus which provides a small oscillating drive field. Whereas VSM oscillates a statically-magnetized sample, ACMS oscillates a field across a static sample and is well suited to the study of magnetic phase transitions as it probes $\chi'$ directly. The diverging magnetic susceptibility associated with a transition into long range order is then directly observed by the pickup coil response. For fields oriented along $z$, two peaks are clearly resolved, near the same temperatures where sharp peaks are observed in the heat capacity measurements (see Fig.~\ref{fig:magnetism}(c)). The ability for ACMS to distinguish closely spaced transitions can be contrast against the modest kink observed for this same sample in this orientation via the VSM plot in Fig.~\ref{fig:simple}(d). However, as we increase the frequency of the 3~mT AC drive field, the higher temperature transition is observed to shift to even higher temperatures -- an unexpected response for a transition into long-range order. The extracted peak temperatures $T_\text{\rm pk}$ are plotted in the inset of Fig.~\ref{fig:magnetism}(a) for each drive frequency.

While heat capacity indicates two sharp transitions at 10.3~K and 9.5~K respectively (See Appendix Fig.~\ref{fig:heat_capacity}), there is an undeniable frequency dependent peak in the interval $10.3~\text{K} < T_\text{\rm pk}(\omega)< 11~\text{K}$ in the ACMS measurements, and the peak temperature T$_\text{pk}$ fits well to the Vogel-Fulcher law used to describe canonical spin glasses like Cu$_x$Mn and Eu$_{1-x}$Sr$_x$S \cite{Tholence_1981,Maletta_1979}. The Vogel-Fulcher frequency $f_\text{AC}$ has an exponential dependence on the activation energy $E_\text{a}$ over the difference of the temperature $T_\text{pk}$ (where the susceptibility peaks) and characteristic temperature $T_0$;
\begin{equation}
f_\text{AC} \propto \exp{\left[-E_\text{a}/(T_\text{pk}-T_0)\right]}.
\end{equation}
The $T_0$ extracted from the Vogel-Fulcher fit is the asymptotic DC limit of the glass transition, fitted here to $T_0=10.3\pm0.1$~K. This matches exceptionally well to the location of the zero-field peak in the heat capacity measured to be $T_\text{N1}=10.34$~K. $T_0$ is often interpreted as relating to an incipient long-range order which is destabilized in the cluster-glass phase \cite{Mydosh}. The agreement between the heat capacity peak and $T_0$ is remarkable, as canonical spin glasses do not show a sharp peak in heat capacity at their freezing temperatures, but rather a broad hump above $T_0$. This suggests that the application of a very modest oscillatory field perturbs the system away from a long-range ordered phase at 10.3K, suggesting that it may instead support a more complex spin texture in the same temperature range. This could be a cluster-glass state as proposed in the related system Eu$_5$In$_2$As$_6$, or some underlying frustration of non-collinear spin texture. 

The influence of the oscillating field is made even more apparent when we rotate its direction. In Fig.~\ref{fig:magnetism}(b), we demonstrate the effect of applying the AC drive field along $H\parallel y$. For AC field oriented off the $z$ axis, a sharp frequency-independent peak is manifest at 10.3~K -- the transition is consistent with the heat capacity result. This striking direction-dependence in the presence of a small perturbative field ($\mu_0 H_\text{AC} = 3$~mT) suggests that the competition between long-range order and the cluster-glass phase is incredibly tight. While long-range order appears to win out at low temperature in all cases, there remains close competition near the transition temperature, resulting in this very unusual behaviour. 

While the frequency response above calls into question whether we indeed have long range order, this claim is made definitive through measurement of isothermal relaxation curves, plotted in Fig.~\ref{fig:magnetism}(d). The sample is cooled from 50~K to the specified final temperature in a 0.5~T field. After thermalizing for 5~minutes, the field is turned off and the magnetization is recorded as a function of time. For an ordered system with a static moment, one expects the magnetization to endure indefinitely. Similarly, a paramagnetic system should also show zero relaxation after the field is turned off, as the magnetization should follow the field to zero instantaneously. In Fig.~\ref{fig:magnetism}(d), we observe such a time independent profile only for temperatures $T < 9$~K and $T> 11$~K, whereas the magnetization shows a logarithmic decay for temperatures within the interval $9.5~\text{K}  < T < 10.5$~K. This coincides with the interval between $T_\text{\rm pk}$ and T$_\text{N2}$ from ACMS. We conclude from this that an applied field $H\parallel z$ destabilizes the long-range ordered transition to a canted AFM at $T_\text{N1}=10.3$~K, favouring instead the freezing of a cluster-glass state in the same range of temperatures. At further lower temperatures however, regardless of the presence of the applied field, long range order wins with a second transition at $T_\text{N2}=9.6$~K. This is internally consistent with the local maxima in the saturation moment plot in Fig.~\ref{fig:phase}(b), which demonstrates that this canted AFM phase below $T_\text{N2}$ is distinct and competitive with that at $T_\text{N1}$. 

\subsection{Interplay of Spin Correlations and Colossal Magnetoresistance}

\begin{figure}
    \centering
    \includegraphics[width=\columnwidth]{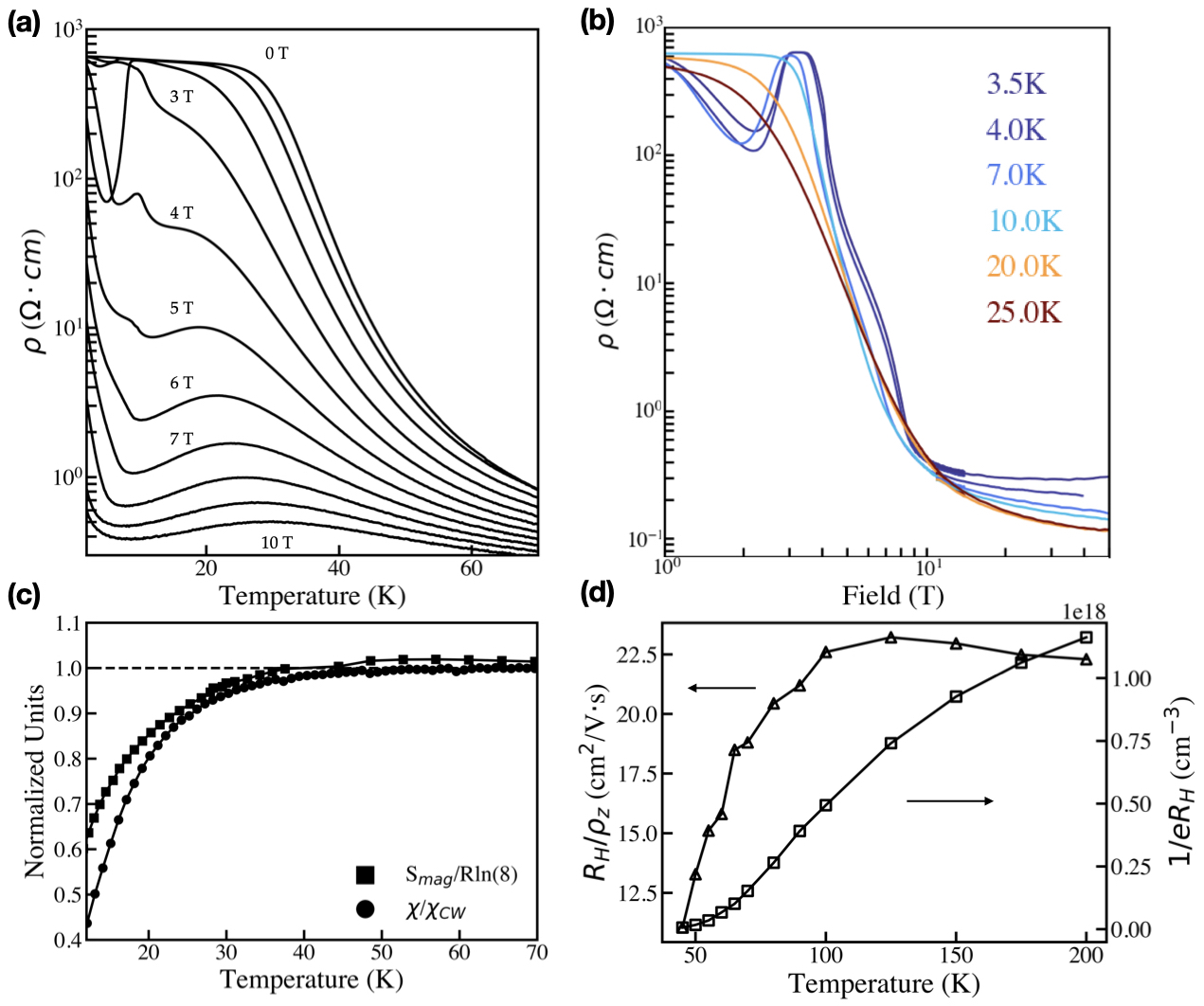}
    \caption{Negative colossal magnetoresistance for $H\parallel y$ is plot in (a), with angle-dependence of $H\in \Pi_{xy}$ at fixed temperatures is provided in the inset. Similar data plot instead for fixed temperatures out to 40~T are shown in (b), with dashed lines indicating data acquired in pulsed fields, whereas solid lines were collected in DC field.
    Magnetic entropy as a fraction of free $J = 7/2$ moments is plot alongside the magnetic susceptibility renormalized by its own expected value for free moments is plotted in (c).
     In (d), the carrier mobility (left) and concentration (right) are plot as a function of temperature, as extracted from fits to the slope of the Hall voltage $V_{xy}$ in the low field limit ($\mu_0 H<1$~T). }
    \label{fig:MR}
\end{figure}

The effects of quenched spin disorder in the system are not restricted to the temperature interval between $T_\text{N1}$ and $T_\text{N2}$, but rather manifest at intermediate and low temperature scales affecting nearly all experimental probes we have applied to the material. The persistent effects of high-energy Eu-Eu exchange interactions are most directly observed through consideration of magnetic heat capacity upon subtracting a phonon background (for details of measurement see Appendix). Given the material's insulating nature, the remaining heat capacity is associated with the magnetic degrees of freedom. We define
\begin{equation}
S_\text{mag}(T) = \int_{T_\text{min}}^{T} dT' \frac{C_\text{mag}(T')}{T'}
\end{equation}
where $T_\text{min} = 0.35$~K is the base temperature of the cryostat. In the limit that the $J=7/2$ moments arising from Eu$^{2+}$ ions are purely independent (i.e. $k_\text{B} T \gg J_{ij}$), the magnetic entropy $S_\text{mag} = R\mathrm{ln}(8)$. In Fig.~\ref{fig:MR}, the computed magnetic entropy is plotted from $10-60$~K. Rather than jump to $R\mathrm{ln}(8)$ at $T_\text{N1}$ as one may expect in moving from a purely paramagnetic to canted-AFM state, the full magnetic entropy is recovered only very slowly, with the transitions manifesting simply as a modest change in slope. 

Only at temperatures in the range $50$-$60$~K do we recover the full magnetic entropy of the Eu$^{2+}$ lattice, indicating that some level of local correlations between Eu moments persist far above the ordering temperature. While this gives evidence to significant magnetic frustration in the system, such a conclusion would appear to run counter to the conventional estimation of the exchange frustration parameter $f =T_\text{N}/T_\text{CW} = 10.3/5.5 = 1.87 > 1$, which would itself naively appear to suggest a lack of frustration. However, it is essential to note that the simple frustration parameter $f$ is an appropriate proxy for frustration when a single exchange parameter is dominant \cite{Ramirez_1994}. Only in this limit is there a straightforward correspondence between $J_{ij}$ and $T_\text{CW}$, the latter of which should be thought of more as a weighted sum over all exchange terms in the system. Given the rather complex exchange network illustrated in Appendix Fig.~\ref{app_fig:arrhenius}, the long tail of the magnetic entropy is a more suitable metric for frustration in the present system. 

Our interpretation of the magnetic entropy lines up very well with the evolution of the magnetic susceptibility, as plotted in Fig.~\ref{fig:MR}(c) as $\chi/\chi_\text{CW}$. Similar to $S_\text{mag}$, at high temperature we expect the susceptibility to reflect that of independent $J=7/2$ moments. Instead, we observe the onset of correlations amongst Eu$^{2+}$ moments at effectively the same temperatures where the magnetic entropy begins to reduce from $\mathrm{R}\mathrm{ln}(8)$. This would be expected at the onset of short range correlations or magnetic clusters of Eu$^{2+}$ moments well above T$_\text{N1}$, and validates this picture. 

In the absence of long range order, the information we can glean from susceptibility and heat capacity in this intermediate cluster regime is limited. To what extent can we further elucidate this phase? To do so, we turn to magnetoresistance, where we observe strong indications of correlations between thermally activated carriers, and the local Eu moments, far above the ordering temperature. As plotted in Fig.~\ref{fig:MR}(a), the magnetoresistance becomes significant at lower temperatures, reaching an order of magnitude reduction of resistivity $\rho_{zz}(T)$ by 50~K and $\mu_0 H=10$~T, and an MR of $600,000\%$ at 25~K. As evidenced by the high field data in Fig.~\ref{fig:MR}(b), CMR begins to saturate above 40~T and the material does not become metallic out to the largest fields measured. This suggests that the relevant mechanism for CMR is associated with spin-scattering as opposed to a field-driven closure of the band gap. This is consistent with our expectation from our Arrhenius fitting of a $35$~meV bandgap from the temperature dependent resistivity which, using the expression $E = g_j\mu_B m_s B$ while assuming a Land\'e g-factor of $2$ and polarized Eu$^{2+}$ spin $m_s = 7/2$, is equivalent to $7800$~T.  Similar to the temperature evolution, angle-dependent magnetoresistance (AMR) begins to reveal onset of non-trivial magnetic correlations around 60~K, as the magnetic anisotropy of the system's low-temperature phase diagram begins to manifest. This is evident from the AMR plot in Fig. \ref{fig:anisotropy} of the Appendix.

The Hall effect in principle allows us to assess the 
evolution of the carrier concentration and mobility with temperature and can shed light on the effect these magnetic clusters have on the carrier density and mobility.
However, as we discuss below, the longitudinal and Hall resistivity data as a function of field cannot be modelled by a single or two-carrier model, making it difficult to assess the effect of temperature on the carrier's mobility.  Instead, we track the zero field Hall number as well as its ratio to the longitudinal resistivity as a function of temperature, as plotted in Fig.~\ref{fig:MR}(d).
The asymptotic decay of the zero field Hall number towards zero at low temperature indicates that the already modest carriers present in the system at high temperature are thermally activated, consistent with our Arrhenius-fitting of the magnetoresistance (Appendix Fig.~\ref{app_fig:arrhenius}). 

The ratio of Hall number to longitudinal resistivity sharply declines around 70~K, not far above the same temperatures where we begin to observe bulk indications of short-range correlations/magnetic cluster formation from heat capacity and magnetic susceptibility.  Interpreting this ratio as mobility, this collapse can be attributed to scattering of carriers from the magnetic fluctuations, likely coming from magnetic clusters which continue to grow as temperature is reduced. Put another way, scattering of dilute carriers by magnetically ordered clusters gives rise to formation of magnetic polarons, which supersede the role of free carriers in the system with an enhanced effective mass. As clusters continue to grow, propagation of a carrier through the disordered magnetic landscape requires a series of spin-flip events, resulting here in the decrease in the Hall ratio. 

Magnetic polarons have previously been seen in a variety of narrow-gap semiconductors such as the manganites \cite{kusters_magnetoresistance_1989,moshnyaga_2009}, and other Eu-based compounds such as EuB$_6$ \cite{das_2012}.  Of particular relevance is Eu$_5$In$_2$Sb$_6$ which exhibits the same structure as \EuSnAs{} and has recently been reported \cite{rosa_colossal_2020,crivillero_surface_2022,ghosh_colossal_2022,souza_microscopic_2022} to host magnetic polarons.  In this system, magnetic polarons have been shown to increase in size as temperature is lowered below $200$~K while the resistivity rises and CMR begins to develop.    While this picture is consistent with our data on \EuSnAs{}, further measurements are required to prove the role of magnetic polarons are the same as suggested by studies in Eu$_5$In$_2$Sb$_6$.


\section{Discussion}
\label{sec:discuss}
As shown in Fig.~\ref{fig:colormap}, $d\rho/dT(H,T)$ provides a direct correspondence to the magnetic phase diagram measured via both heat capacity and magnetization. This correlation is found across all crystal orientations of the material, reflecting the direct relationship between the magnetoresistance and the magnetic order in the material. (Signatures of this correspondence can also be identified in continuous angle-dependent magnetoresistance curves.) The richness of the magnetic phase diagram of \EuSnAs{} reflects the underlying complexity of the exchange interactions, and it is reasonable that this is the cause of the CMR. The picture of current carrying magnetic polarons is consistent with our observations and already discussed for the related system Eu$_5$In$_2$Sb$_6$. \cite{crivillero_surface_2022, Rosa_2022, rosa_colossal_2020, souza_microscopic_2022} Here the dressed quasiparticles are increasingly localised by the growing influence of the exchange interactions, a process that is rapidly suppressed by the application of magnetic fields. We note that other models could also explain the CMR observed here. Recent studies on the magnetoresistance of non-collinear magnets thought to host loop-currents, like Mn$_3$Si$_2$Te$_6$, \cite{zhang_control_2022} show very similar CMR to that reported here. Further studies are required to distinguish these possibilities.


While the CMR reported here is exceptionally large, we believe that it is in fact only an approximate lower bound on the bulk magnetoresistance in this material, as the low temperature zero field resistivity appears to shunt across a low-resistance path. This manifests as a kink or knee in $\rho_{ii}$(T) below approximately 30K in low field. This limits our ability to evaluate the absolute scale of negative CMR in this system. As articulated in Fig.~\ref{fig:colormap} above, the low temperature regime is characterized by a wealth of fine structure in the magnetoresistance, consistent with the phase boundaries identified in heat capacity and magnetic susceptibility. In low field in particular, we see an intermediate phase where $\rho_{zz}$ drops by an order of magnitude between $T_\text{N1}$ and 5~K, before rising again at the lowest temperatures measured.


The simplest approach to understanding the magnetoresistance of a narrow gap semiconductor is by modelling the system as a two carrier model with the carrier's mobility and density taken to be independent of field. In this simple picture, the conductivity $\sigma$ should scale as negative $B^2$ and a resistivity $\rho$ which scales as positive $B^2$.  However, the magnetoresistance of \EuSnAs{} has a negative $B^2$ dependence, making any multi-carrier model with field independent of mobility and carrier density impossible to fit.
This implies that the carrier mobility or density evolves as a function of field and a simple multi-carrier model is insufficient to describe the physics in this material. This is consistent with the interpretation of a field-dependent magnetic order and the presence of meta-magnetic transitions observed above. 

The apparent plateau in the resistivity at low field and low temperature indicates that the current appears to shunt across a higher conductivity pathway.   Surface states would provide a plausible explanation for this observation, as their comparatively small current capacity would limit our ability to identify their contribution at high temperatures where the large cross section of the samples (on the order of \SI{0.04}{mm^2}) would dominate the transport. Moreover, the absence of any associated features in the bulk characterization of susceptibility and heat capacity in this same region of phase space cast doubt on a bulk transition here. This is similarly true for the low temperature magnetoresistance as the bulk CMR pulls $\rho_{zz}$ below the shunt value (for this sample $\rho_{zz}  < 70$~\unit{\ohm\cm}). 

A surface state contribution to the resistivity in \EuSnAs{} is reasonable. Topologically trivial surface states in narrow gap semiconductors have been investigated in narrow gap semiconductors such as InSb and (Hg,Cd)Te \cite{sze_semiconductor_2012}.  In this case, dangling bond states caused by surface termination in semiconductors induce bandbending at the surface and results in a surface accumulation layer of carriers. Topological surface states may also contribute as predicted in DFT calculations for materials in the Eu$_5$M$_2$X$_6$ family, assuming an A-Type AFM magnetic order \cite{Varnava_2022, Rosa_2020}.  However, the properties of \EuSnAs{} measured in this study point toward a magnetic structure that is far more complex than an A-Type AFM, making any topological classification assuming this structure hard to trust. To date, no experimental evidence has been found that these materials are indeed topological. 


\section{Conclusion}
Magnetic Zintl semiconductors offer a wealth of opportunities for the development and study of new phenomena through their combination of narrow gaps and geometric confinement. We have presented a comprehensive characterization of the magnetic and electronic properties of \EuSnAs{} across its entire low-temperature phase diagram. The presence of multiple symmetry-distinct Eu sites per unit cell, combined with the exchange-frustrated network of magnetic interactions in the system provide the requisite conditions to support a complex magnetic phase diagram, which itself shows a very tight competition with cluster-glass dynamics upon application of even very modest applied fields. This tension between long-range order and glass physics manifests in all experimental probes surveyed, including magnetization, heat capacity, and magnetoelectrical transport. This enables observation of Eu-exchange interactions for temperatures almost an order of magnitude greater than the ordering temperature, a fact which is surprising given the absence of sufficient carriers to mediate exchange via for example the RKKY mechanism. Most notably, the interaction between the dilute thermally activated carriers in the system and the local moments give rise to colossal magnetoresistance of at least 600 000$\%$, likely through the formation of magnetic polarons via scattering of carriers off of magnetic clusters. At the lowest temperatures measured, the bulk conductivity is shunted via what appears to be surface states, 
Future experiments will be required to elucidate the true nature of the various metamagnetic orders in the system as well as the nature of the shunting of the electrical resistivity at low temperature, developments which will be essential to further understanding of the topology in this system. 
 
\begin{acknowledgments}
This research was supported by he Air Force Office of Scientific Research (AFOSR) Multidisciplinary University Research Initiative (MURI) grant on TOPological Flat bands fOr Correlated Electron systems (TOPFORCE) under the award number FA9550-22-1-0270. R. P. D. was supported by the Canadian Government under a Banting Fellowship. Work at the
National High Magnetic Field Laboratory was supported by
NSF Cooperative Agreements No. DMR-1644779 and No.
DMR-2128556, the U.S. Department of Energy (DOE), and
the State of Florida. J.S. acknowledges support from the DOE
Basic Energy Sciences FWP “Science of 100 T.”

\end{acknowledgments}

\bibliography{Eu526_CMR}

\section{Appendix}
\subsection{Eu-Eu Exchange Pathways}
\begin{figure}[H]
    \centering
    \includegraphics[width=0.5\columnwidth]{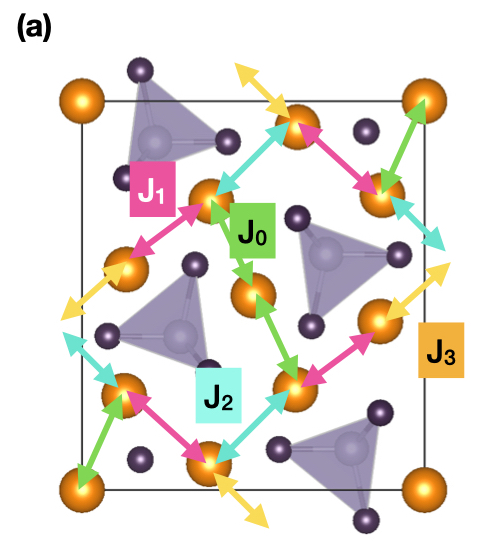}
    \caption{In panel (a), the $x-y$ plane of Eu is drawn alongside schematic illustration of nearest-neighbour exchange pathways between inequivalent Eu moments. }
    \label{app_fig:unitcell}
\end{figure}

The unit cell of Eu$_5$Sn$_2$As$_6$ is complex, with three separate Europium Wyckoff positions and various Eu-Eu distances even within the ab-plane, shown in Fig. \ref{app_fig:unitcell}.  This allows for a competition between Heisenberg exchange mechanisms and allows for a competition of magnetic orders within this material.

\subsection{Arrhenius Model Fit of Electric Resistivity}
\begin{figure}
    \centering
    \includegraphics[width=\columnwidth]{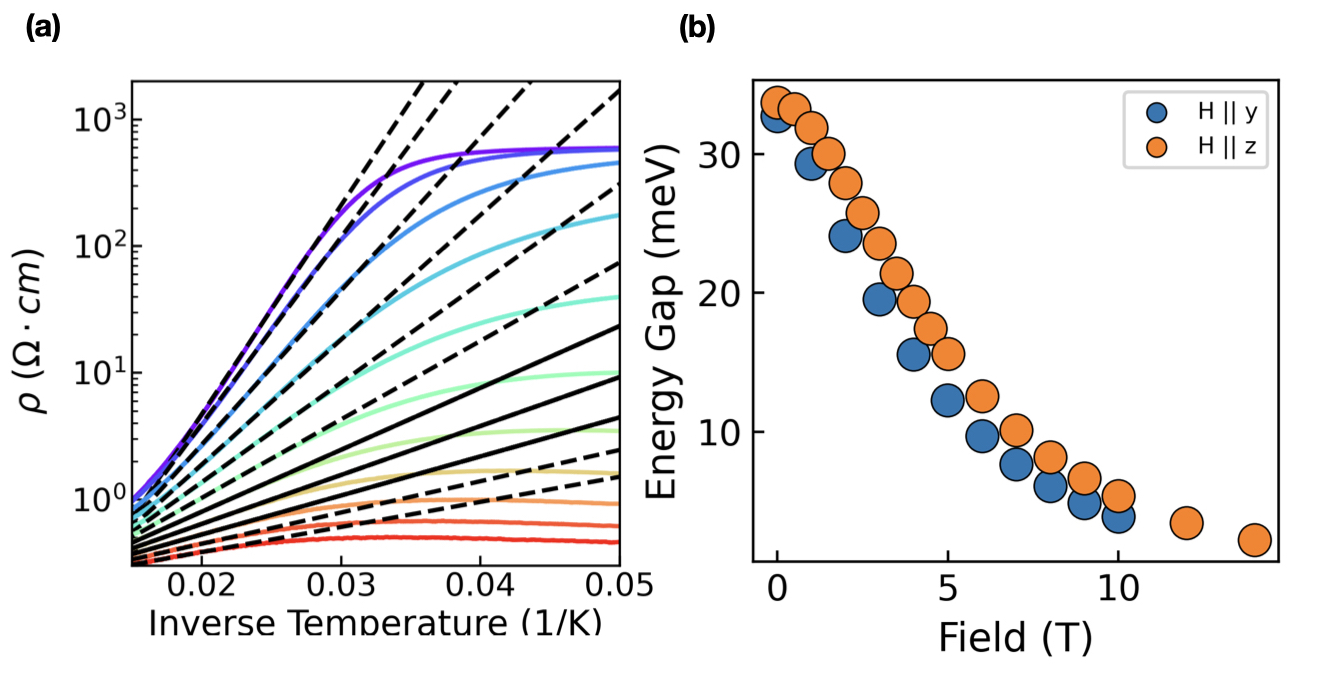}
    \caption{(a) Electrical resistance of Eu$_5$Sn$_2$As$_6$ at different fields from 0T to 10T plotted in a log scale against inverse temperature  The gap $\Delta$ is plotted as a function of magnetic field for two orientations, H$\parallel$y and H$\parallel$z.  }
    \label{app_fig:arrhenius}
\end{figure}

\begin{figure}
    \centering
    \includegraphics[width=\columnwidth]{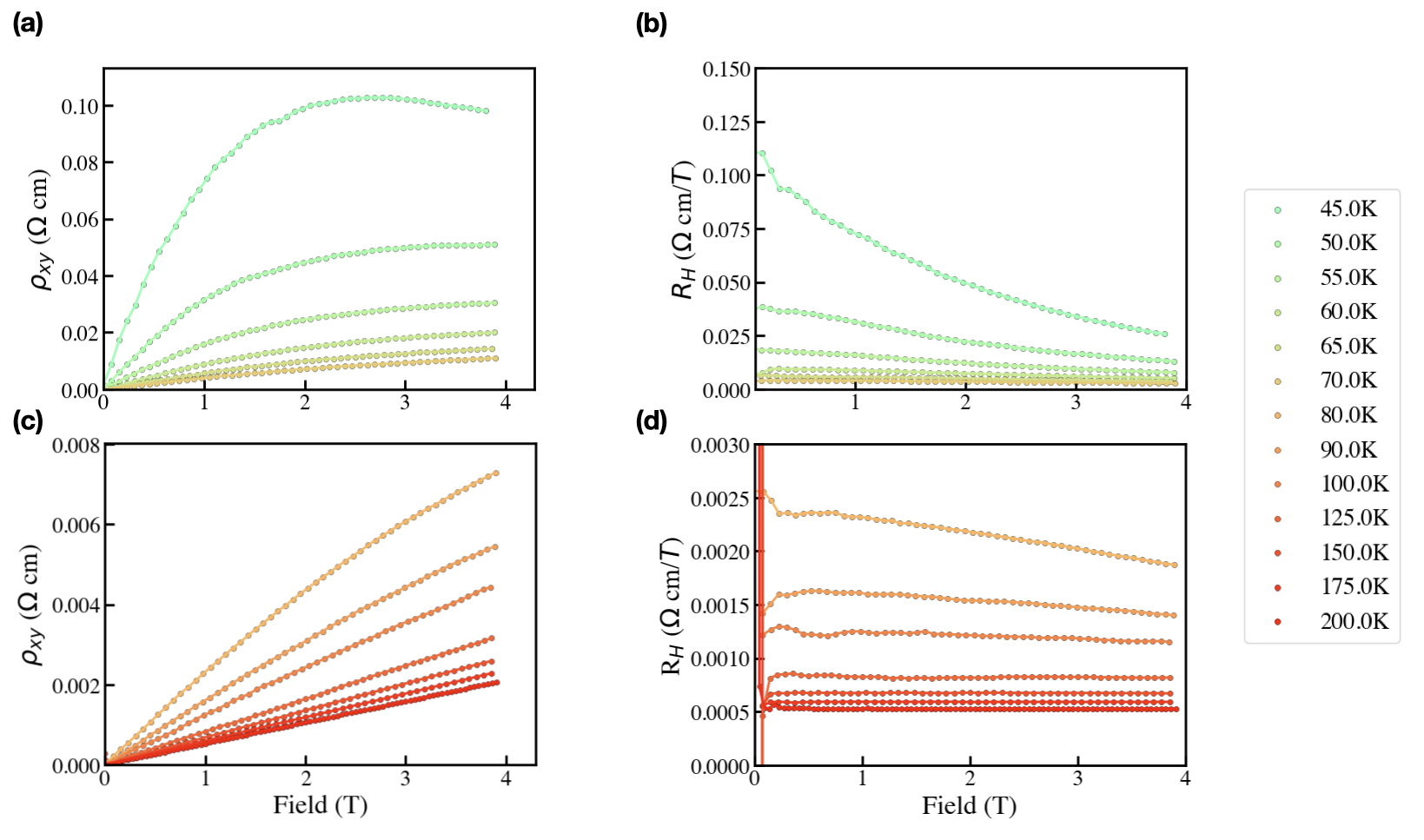}
    \caption{The Hall Resistivity $\rho_{xy}$ is plotted for temperatures ranging from $T\geq 45$~K in (a) and $T\geq80$~K in (c).  The Hall number for higher temperatures $50\leq T \leq 80$ is plotted in (b) while $T\geq 80$ is plotted in (d) as a function of field.}
    \label{app_fig:hall}
\end{figure}

The electrical resistance as a function of temperature is fit to the Arrhenius model of thermally activated carriers in \ref{app_fig:arrhenius}, indicating a gap of $35$~meV at 0 Temperature and decreasing with applied field.  Two separate field directions are shown.
\begin{equation}
\rho\sim \exp{(-\Delta/k_BT)}
\end{equation}
\subsection{Hall Number and Hall Resistivity}
The Hall Number is plotted as a function of Field and shown to be linear at high temperature but below 60K, begin to curve and saturate.  

Hall Effect measurements below $\sim50$~K are dominated by longitudinal resistivity as any amount of misalignment in the six terminal Hall terminal measurement dominates the signal.  While the standard method of removing misalignment voltage is by antisymmetrizing with respect to field, the appearance of CMR significantly decreases the ratio of the Hall to longitudinal resistivity.  In other words, as temperature decreases, the Hall Resistance does not grow in proportion to the longitudinal resistivity and below $\sim40$~K, the signal-to-noise ratio determined by our lock-in setup is insufficient to detect the Hall voltage even after antisymmetrizing.

\subsection{Hall Number and Hall Resistivity}
\begin{figure}
    \centering
    \includegraphics[width=\columnwidth]{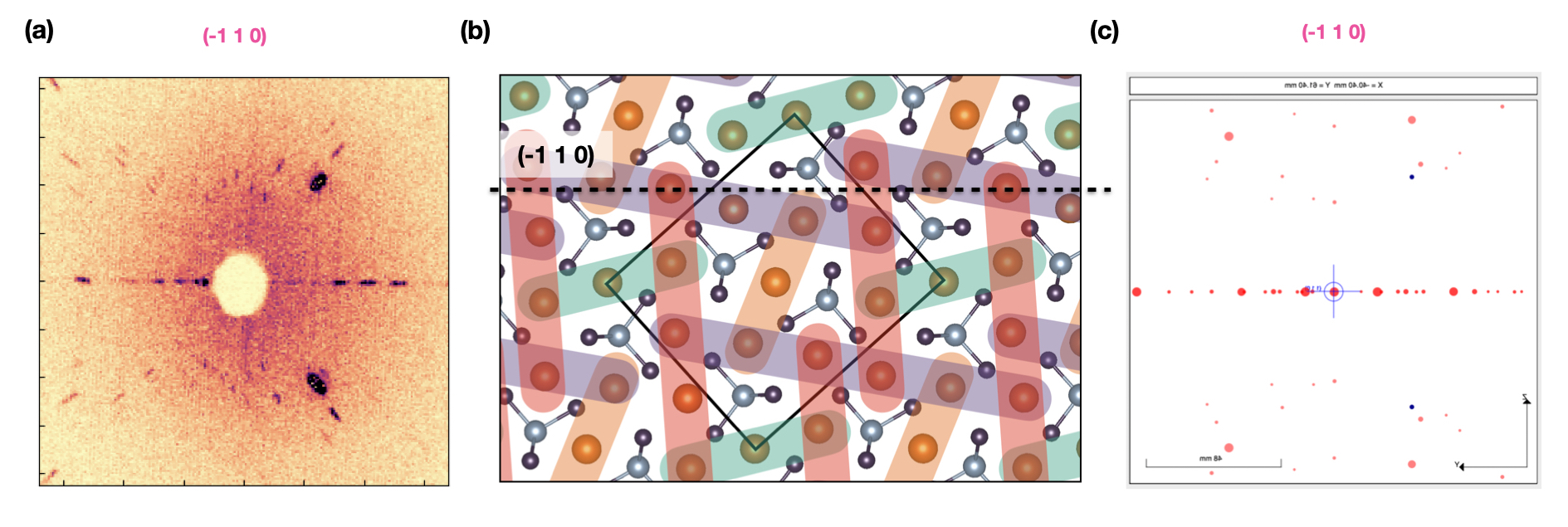}
        \caption{(a) Laue Diffraction on Eu$_5$Sn$_2$As$_6$.  (b) Neighbouring Eu atoms
    form staggered placquettes of variable extent, illustrated by
    the shaded regions. The natural crystal termination along
    the (-1 1 0) plane is indicated by a dotted line. The QLaue fit from the Eu5Sn2As6 cif file is plotted in (c).}
\label{fig:Laue}
\end{figure}
The Eu$_5$Sn$_2$As$_6$ crystals come out as prisms with a cross-section that has one large crystal facet.  We orient ourselves so that the longest side of the needle-like crystals is the $z$ direction while the largest crystal facet is perpendicular to the $x$ direction. 
Laue diffraction Analysis using the software program QLaue suggests that this $\hat x$ is parallel to the (-110) direction, cutting the unit cell as seen in Appendix Fig.~\ref{fig:Laue}(b).  

\begin{figure}
    \centering
    \includegraphics[width=\columnwidth]{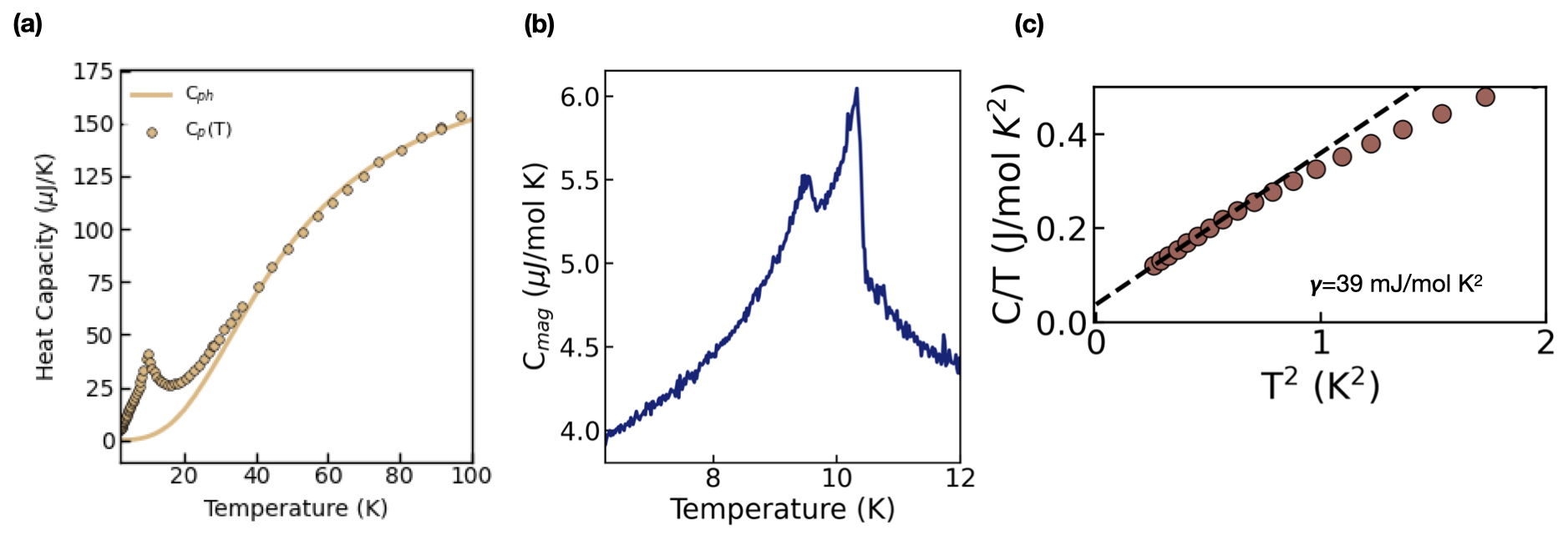}
    \caption{The phonon contribution to the zero field Heat Capacity of Eu$_5$Sn$_2$As$_6$ is fit to a Debye curve in plot (a) while the phonon-subtracted heat capacity is shown in (b).  }
    \label{fig:heat_capacity}
\end{figure} 
The total heat capacity is broken up into contributions from the electronic, vibrational (phononic), and magnetic degrees of freedom.
\begin{equation}
C_{raw}(T) = C_{mag}(T) + C_{ph}(T) + C_{el}(T)
\end{equation}
We fit the Heat Capacity $C_{ph}(T)$ as a function of temperature $T$ to the Debye equation, 
\begin{equation}
C_{ph}(T) = 9Nmk_B\left(\frac T {\theta_D}\right)^3 \int_0^{\theta_D/T} x^4 \frac{e^x}{(e^x-1)^2}\text{ d} x
\end{equation}
where $m$ is the mass of the sample, $N$ is the number of atoms in the sample, and $k_B$is Boltzmann's constant.  As the phonons tend to dominate the heat capacity at higher temperatures, this fit is done in the temperature range 60-80K to $\theta_D = 196$ K and is shown in Appendix Fig.~\ref{fig:heat_capacity}.  The electronic contribution to the heat capacity is found by fitting the Sommerfeld model, $C_{el}(T) = \gamma T$ with Sommerfeld coefficient $\gamma = 39$ mJ/mol K.

Finally, the rest of the heat capacity is attributed to the magnetic degrees of freedom as the magnetic heat capacity, shown in Appendix Fig.~\ref{fig:heat_capacity}. This is integrated over the magnetic transition to get the magnetic entropy $S_{mag}$ as, the deviation of the magnetic entropy from $S = R \ln J(J+1) =R \ln 8$ for Eu$^{2+}$ ions with $J=S=7/2$ is plotted in Fig. 5 of the main text.


\begin{figure}
\centering
    \includegraphics[width=\columnwidth]{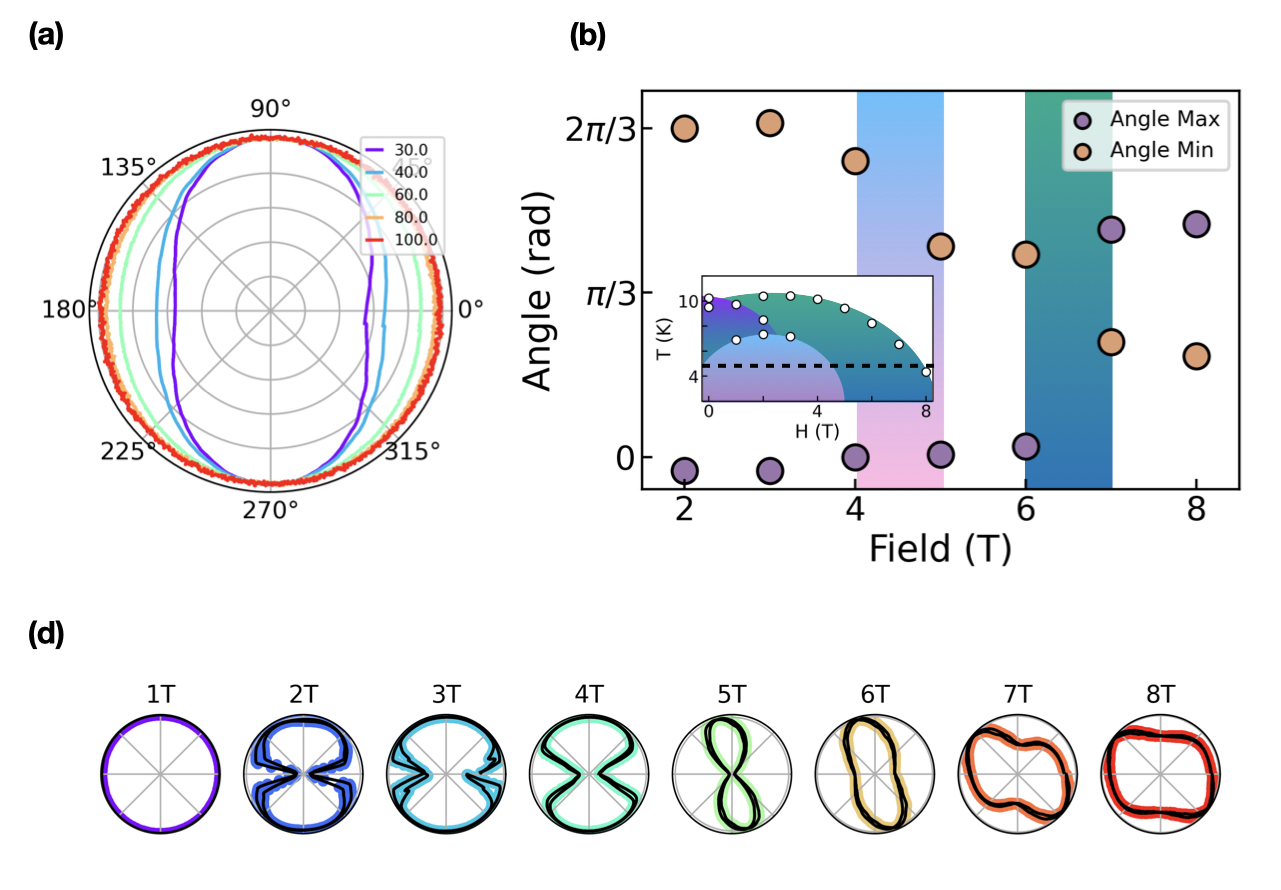}
\caption{In (a), AMR with applied field 2T is shown at different differentures.
    The rotation of the extrema line up with the phase diagram extracted from heat capacity, as shown in panel (b).   Evolution of the AMR in the $\Pi_{xy}$ at T=5K is shown for different field strengths in (c) indicating a change in ab-plane scattering symmetry as a function of field.}
    \label{fig:anisotropy}
\end{figure}
The Angular Magnetoresistance (AMR) is done in the standard four terminal electric transport configuration with the applied current along the z direction and the rotation in the ab-plane with $0^\circ$ aligned to the $x$-axis.  The data as a function of temperature in Appendix Fig.~\ref{fig:anisotropy}(a) indicates the breaking of U(1) symmetry in the ab-plane as temperature decreases below 80K, perhaps in-conjunction with the formation of magnetic polarons.  As plotted in Appendix Fig.~\ref{fig:anisotropy}(b), the angles at which the AMR is extermal at $T = 5$ K seem to evolve concurrently with changes in the the Heat capacity as a function of field at  $T = 5$ K.  Since the heat capacity at low temepratures is correlated to changes in the magnetic susceptibility and resistivity as seen in Fig. 4 in the main text, this suggests that the cause for this anisotropy is linked to changes in the bulk magnetic order.  
\subsection{Microwave Impedance Microscopy}
\begin{figure}
\centering
    \includegraphics[width=\columnwidth]{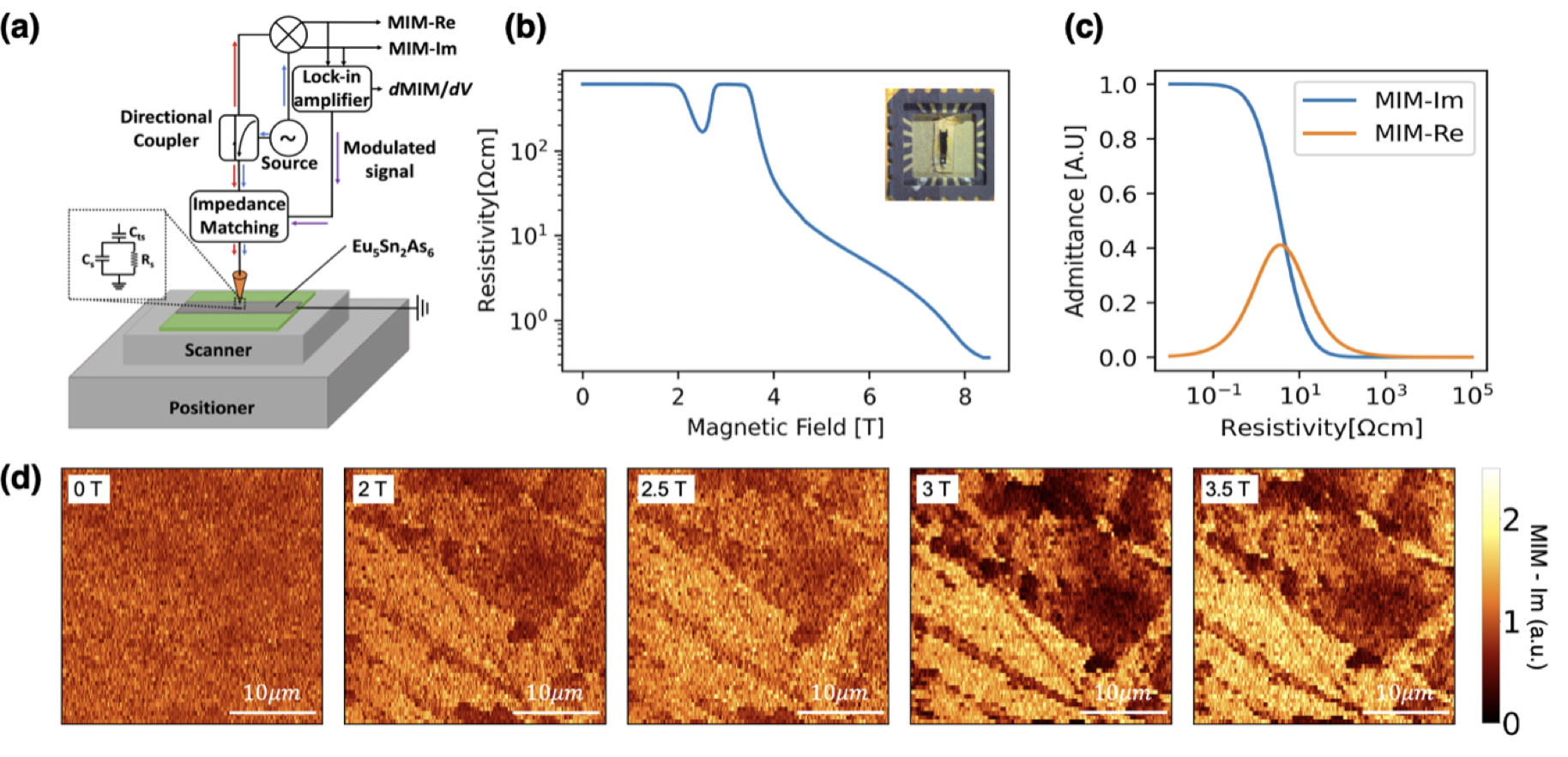}
    \caption{\textbf{Spatially-resolved conductivity profile of Eu5Sn2As6 obtained using microwave impedance microscopy (MIM).} (a) Schematic illustration of the MIM measurement setup, where the tip-sample interaction with the sample is abstracted using a lumped-element model. (b) Two-terminal transport measurement conducted in as a function of out-of-plane magnetic field ($H\parallel y$) and at $T = 4$~K. Inset: Optical image of the Eu$_5$Sn$_2$As$_6$ crystals crystal. (c) Theoretical MIM response curves, simulated using finite element analysis at 3 GHz, show how the reflected microwave signal depends on the local resistivity of the sample. The real and imaginary parts of the reflected signal are denoted MIM-Re (red) and MIM-Im (blue), respectively. (d) Spatially-resolved MIM images obtained in at a series of increasing magnetic fields, applied out-of-plane. MIM-Im, the imaginary part of the microwave response, is plotted in each image. The MIM measurement frequency is 3 GHz, and the temperature is 100 mK. Each scale bar is 10 $\mu$m.}
    \label{fig:MIM}
\end{figure}

We performed microwave impedance microscopy (MIM) measurements to image the evolution of the real-space conductivity profile as a function of the out-of-plane magnetic field ($H\parallel y$) at low temperatures (T = 100~mK). As illustrated in Fig. 8, we observed microscopic spatial variations in the amplitude of the reflected MIM signal, which suggests an inhomogeneous conductivity landscape with circular pockets of high resistivity (dark clusters) that coexist with large-area conductive domains (bright rectangular regions). Further imaging work would be required to investigate the percolation of magnetic polarons in detail.

\end{document}